\documentclass{pasj00}
\draft

\begin{document}
\SetRunningHead{Nakagawa et al.}{AKARI Cryogenics}
\Received{2007/05/31}
\Accepted{2007/08/09}

\title{Flight Performance of the AKARI Cryogenic System}

\author{%
Takao~\textsc{Nakagawa},\altaffilmark{1,2}
Keigo~\textsc{Enya},\altaffilmark{1}
Masayuki~\textsc{Hirabayashi},\altaffilmark{3}
Hidehiro~\textsc{Kaneda},\altaffilmark{1}
Tsuneo~\textsc{Kii},\altaffilmark{1}
Yoshiyuki~\textsc{Kimura},\altaffilmark{3}
Toshio~\textsc{Matsumoto},\altaffilmark{1}
Hiroshi~\textsc{Murakami},\altaffilmark{1}
Masahide~\textsc{Murakami},\altaffilmark{4}
Katsuhiro~\textsc{Narasaki},\altaffilmark{3}
Masanao~\textsc{Narita},\altaffilmark{1}
Akira~\textsc{Ohnishi},\altaffilmark{1}
Shoji~\textsc{Tsunematsu},\altaffilmark{3}
and 
Seiji~\textsc{Yoshida}\altaffilmark{3}}

\altaffiltext{1}{Institute of Space and Astronautical Science, 
Japan Aerospace Exploration Agency, 3-1-1~Yoshinodai, Sagamihara, Kanagawa 228-0024, Japan}
\email{$^2$nakagawa@ir.isas.jaxa.jp}
\altaffiltext{3}{Sumitomo Heavy Industries, Ltd. 5-2 Soubiraki-cho, Niihama, Ehime 791-8588, Japan}
\altaffiltext{4}{University of Tsukuba, 1-1-1 Ten-nodai, Tsukuba, Ibaraki 305-8571, Japan}
%

%


\maketitle

\begin{abstract}
We describe the flight performance of the 
cryogenic system of the infrared astronomical satellite
AKARI, which was successfully launched on 2006 February 21 (UT).  
AKARI carries a 68.5 cm telescope together with
two focal plane instruments,
Infrared Cameras (IRC) and Far Infrared Surveyor (FIS), 
all of which are cooled down to cryogenic temperature to achieve superior sensitivity.
The AKARI cryogenic system is a unique hybrid system, which consists of
cryogen (liquid helium) and mechanical coolers (2-stage Stirling coolers). 
With the help of the mechanical coolers, 
179 L (26.0 kg) of super-fluid liquid helium can keep the instruments 
cryogenically cooled for more than 500 days. 
The on-orbit performance of the AKARI cryogenics is consistent with the design and
pre-flight test, and the boil-off gas flow rate is as small as 0.32 mg/s.
We observed the increase of the major axis of the AKARI orbit, which can be explained by
the thrust due to thermal pressure of vented helium gas.
\end{abstract}

\section{Introduction}

In order to achieve superior sensitivity, infrared 
telescopes in space must be cooled down
to cryogenic temperatures. 
IRAS \citep{neugebauer_84} was the
first infrared astronomical satellite
which demonstrated the effectiveness of 
an infrared astronomical mission with a
cooled telescope. 
IRAS carried about 500 L of liquid helium
to cool its observation instruments, including
a 60 cm telescope.
Its mission life was only 10 months
due to the short hold time of liquid helium.
ISO, which was a European
infrared astronomical satellite with a 60 cm telescope,
carried over 2300 L of liquid helium
to achieve 2.5-year operation life time in space \citep{kessler_96}.
The first Japanese infrared space mission, IRTS \citep{murakami_96},
was not a dedicated satellite but was one of the experimental
instruments onboard the multi-purpose experiment satellite 
SFU (Space Flyer Unit). IRTS was launched in 1995 onboard SFU and the mission life of IRTS 
was about a month,
since the amount of liquid helium onboard IRTS was very small
(90 L).  
In summary, previous missions
required a huge amount of liquid helium to keep
their observation systems cooled for a long time.

The Spitzer Space Telescope (hereafter SPITZER) \citep{werner_04}, which was launched
in 2003, employed a new approach. The telescope
was launched at ambient temperature and was cooled on orbit by the combination of radiative cooling
and helium boil-off vapor. SPITZER utilized an earth-trailing solar orbit.
The major advantage of the orbit is being
away from the heat of the earth.
Hence the outer shell of the cryostat can be cooled very effectively
by radiation and a very long mission life (longer than 5 years) 
is expected with 337 L of liquid helium \citep{werner_04}.

Although the SPITZER's approach works very effectively for 
satellites away from the earth, we cannot apply the same
technique directly to satellites in the near-earth orbit, because the
heat load from the earth is much higher than that for the SPITZER.

We have developed a different type of cryogenic system; 
a hybrid cryogenic system, which consists
of liquid helium and mechanical cryocoolers. 
We use this system for AKARI (formerly known as ASTRO-F),
which is the first Japanese satellite
dedicated for infrared astronomy  \citep{murakami_07}.
the orbit of AKARI is a near-earth orbit, and the
temperature of the outer shell of the cryostat is much
higher than that of SPITZER due to large heat load from the
earth.
The hybrid-cryogenic design enables us to cool the observation system with modest amount 
(179 L) of liquid 
helium for more than a year in the near-earth orbit.
The mechanical cryocoolers also enable observations
at wavelengths shorter than 5~$\mu$m even after
the exhaustion of liquid helium.

AKARI carries two
focal plane instruments (FPI):
Infrared Camera (IRC) \citep{onaka_07} and
Far-Infrared Surveyor (FIS) \citep{kawada_07}.
The AKARI telescope forms a Ritchey-Chretien system
with a primary mirror of 68.5 cm diameter, which is made of
a porous silicon carbide (SiC) core and a 
chemical-vapor-deposited coat of SIC on the surface (\cite{kaneda_05}, \cite{kaneda_07}).
All of these instruments must be cryogenically cooled
to achieve superior sensitivity.

AKARI was successfully launched from 
the Uchinoura Space Center, Japan
on 2006 February 21 (UT) 
and was put into 
a solar synchronous orbit 
above the twilight zone at the
altitude of about 700 km \cite{murakami_07}.

\citet{hirabayashi_07}
discuss details of the design and pre-flight test
of the AKARI cryogenic system.
In this paper, we describe the flight operation and 
the on-orbit performance of the 
AKARI cryogenic system. 


\section{Requirements and Implementation}
\label{sec:req}

In this section, we briefly summarize requirements for the AKARI cryogenics
and their system implementation.

\subsection{Requirements}

Table~\ref{tab:req} summarizes the cooling capability
required for the cryogenic system of AKARI. 
To achieve superior sensitivity, observational instruments
must be cooled down to cryogenic temperature.
We have two aspects on this issue.

\begin{table}
\caption{Required cooling capability of the cryogenic system of AKARI.}
\label{tab:req}
\begin{center}
\begin{tabular}{lll}
\hline
\hline
& Parameter & Values \\
\hline
\multicolumn{3}{l}{With liquid helium} \\
& IRC body temperature & $<$ 7~K \\
& FIS body temperature & $<$ 3~K \\
& Ge:Ga detector temperature & 2.0 - 2.5 K \\
& Stressed Ge:Ga detector temperature & 1.7 K \\
& Detector and heater power directly dissipated & 3.4 mW\footnotemark[$*$]\\
& in the liquid helium & \\
& Telescope temperature & $<$ 7~K \\
& Hold time of Liq. He & $>$ one year \\
\hline
\multicolumn{3}{l}{After the exhaustion of liquid helium} \\
& IRC body temperature & $<$ 35~K \\
& Telescope temperature & $<$ 35~K \\
\hline
    \end{tabular}
  \end{center}
\footnotemark[$*$] This is a peak value, and the average heat dissipation is 1.5 mW.
\end{table}

One aspect is to reduce the thermal background from observational 
instruments lower than natural sky background. 
The temperature requirement for the AKARI's telescope 
in table~\ref{tab:req} is determined from this point of view.
FIS has the most stringent requirement
on this issue \citep{kawada_07}.

The other aspect is to reduce thermal dark current of detectors
low enough so that dark current does not limit the sensitivity.
The temperature requirements for the two focal plane
instruments are determined from this point of view.
Since detector arrays on AKARI have different levels of dark current
as a function of temperature, each instrument has its own
temperature requirement. FIS detectors require the lowest temperature. 

The requirement on the hold time of liquid helium
is determined 
to perform the all-sky survey observation in the infrared.
Ideally, a half-year period of observation can cover the whole sky.
However, due to many practical constraints, such as the moon on the sky
and the south Atlantic
anomaly near the earth, we cannot cover the whole sky in half a year, and
we need at least one year to cover the all sky. Hence 
the hold time of liquid helium is required to be longer than one year.  

One more set of requirements is the temperature of IRC and the telescope 
after liquid helium runs out. This enables 
the near-infrared observations with IRC even after helium runs out only
with mechanical cryocoolers. 

\subsection{Implementation}

\begin{figure}
  \begin{center}
    \FigureFile(100mm,150mm){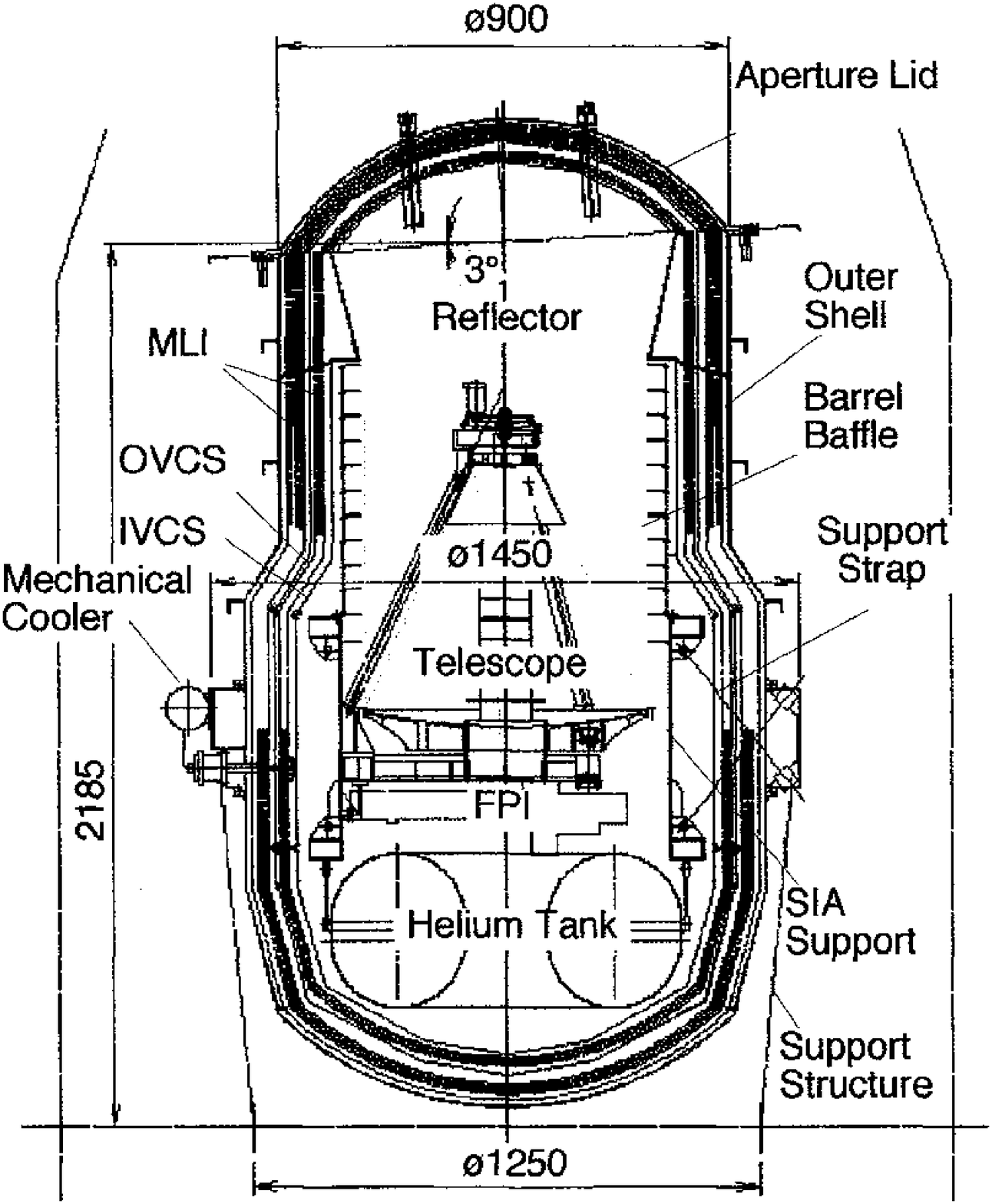}
  \end{center}
	\caption{Cross sectional view of the AKARI cryogenic system.}
	\label{fig:crosscut}
\end{figure}
\begin{figure}
  \begin{center}
    \FigureFile(120mm,150mm){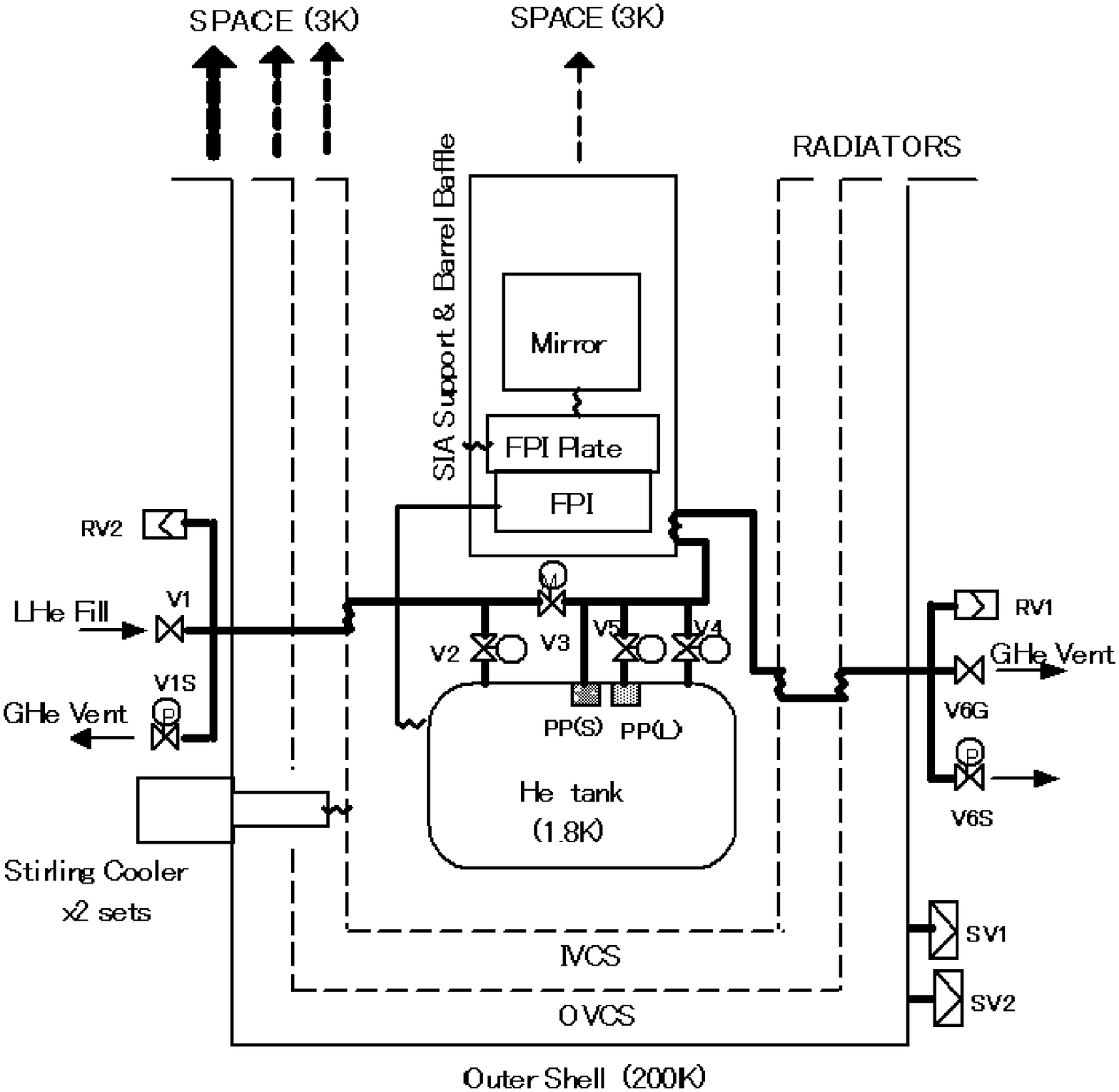}
  \end{center}
  \vspace{-5cm}
  \caption{Conceptual view of the AKARI cryogenic system configuration.}
  \label{fig:overview}
\end{figure}

Table~\ref{tab:spec} summarizes
specifications of the AKARI cryogenics 
implemented to meet
the requirements in table~\ref{tab:req}.
Figure~\ref{fig:crosscut} shows
the cross sectional view of the
AKARI cryostat and 
figure~\ref{fig:overview} illustrates a schematic view of the AKARI
cryogenic system.

\begin{table}
\caption{Specifications of the cryogenic system of AKARI.}
\label{tab:spec}
\begin{center}
\begin{tabular}{lll}
\hline
\hline
Cooling Method & & Hybrid system consisting of cryogen\\ 
	& & and mechanical cryocoolers\\
Cryogen & Type & Super-fluid liquid helium \\
	& Amount & $>$ 25.0 kg at the launch \\
	& Phase separation & Two sets of porous plugs optimized \\
	& & for different flow rates \\ 
Mechanical Cryocoolers & Type & Two-stage Stirling cryocoolers \\
& Number & Two sets \\
& Cooling Power & 200 mW at 20 K per set \\
& Nominal Operation & Simultaneous operation of the two sets \\
& & with half-power operation for each set \\ 
\hline
    \end{tabular}
  \end{center}
\end{table}

The AKARI cryogenic system is a hybrid system of
cryogen (super-fluid liquid helium) and 
mechanical cryocoolers (two-stage Stirling cycle coolers). 
The observational instruments are cooled directly by 
liquid helium ($<$ 2~K) or by boil-off helium vapor ($\sim 4-10$~K).

Two types of porous plugs (figure~\ref{fig:crosscut}) are used
for phase separation of helium.
The two porous plugs
are optimized 
for different flow rates; 5 mg/s for a large porous plug (PP-L) 
and 0.5 mg/s for a small porous plug (PP-S). 
The porous plugs are made of sintered alumina, and their size is 20 mm in diameter
and 5mm in thickness \citep{hirabayashi_07}.

The cryocoolers are implemented not 
to cool observational instruments directly but to reduce the heat
leak to the helium tank, and thereby to extend the hold
time of cryogen on orbit.
Hence the two-stage Stirling coolers are 
thermally connected to the Inner Vapor Cooled Shields (IVCS), 
one of the vapor cooled shields which surround the helium tank
and the observation system, to reduce the heat load to the
helium tank.

In order to reduce the heat
load to the helium tank, 
it is crucial to lower the temperature of the
outer shell of the cryostat.
To protect the cryostat from the direct radiation from the sun,
we put a sun-shield at one side of the cryostat.
We also covered the 
surface of the cryostat with Ag-coated polyetherimide
film (Ag/PEI, \cite{ichino_86}) to reduce the absorption in the visible
light and to increase the emissivity in the infrared, and thereby to
make the radiative cooling effective.

Another key component is the aperture lid (APLD).
The aperture lid is designed 
to keep the vacuum inside the cryostat on the ground
and to be jettisoned on orbit
after the outgas from the satellite drops significantly
and the pressure outside the cryostat approaches
to that of the environmental pressure of gas.
Since the aperture lid blocks some of the radiation paths
from the outer and inner vapor shields, the on-orbit heat load to
the helium tank with the aperture lid is estimated to be about three times larger than that
after the lid is jettisoned. 

\subsection{Two-stage Stirling cooler}

\begin{figure}
  \begin{center}
    \FigureFile(150mm,120mm){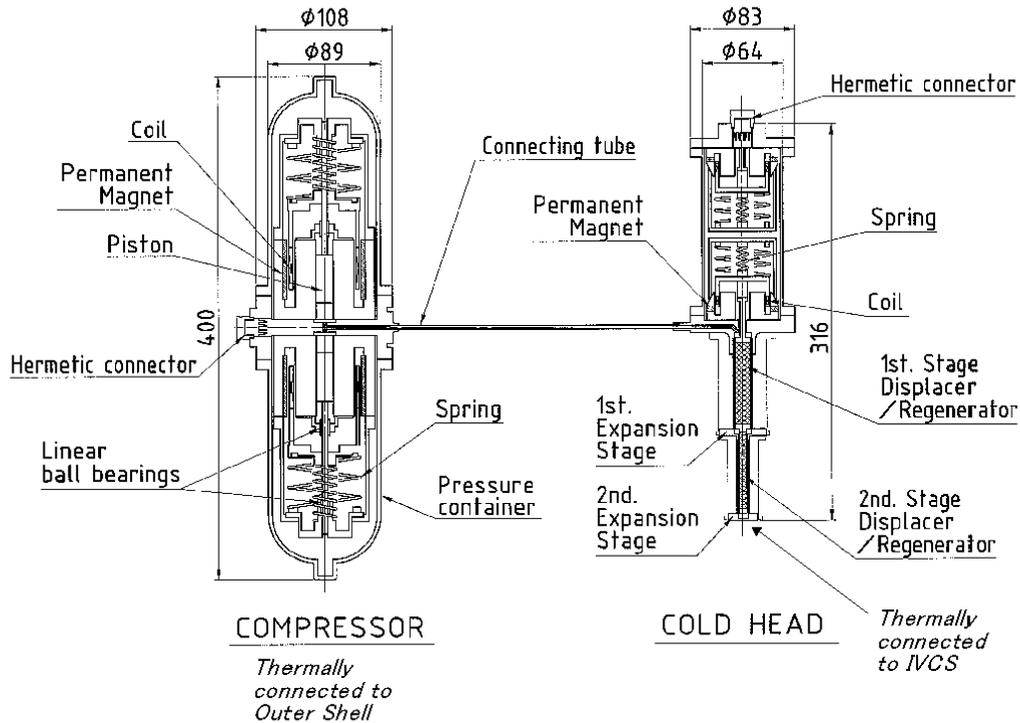}
  \end{center}
  \caption{Cross-sectional view of the two-stage Stirling cryocooler.}
	\label{fig:stirling}
\end{figure}

One of the key features of the AKARI cryogenic system is the
mechanical cryocoolers.
The mechanical coolers adopted for AKARI are
two-stage Stirling coolers \citep{narasaki_04}.
Figure~\ref{fig:stirling} shows a cross-sectional view of the 
two-stage Sirling cooler, which consists of
a compressor, two-stage cold head, and a connecting tube.
Each set of the cryocoolers weighs about 9.4~kg.
Each compressor has two linear pistons operating 
in opposite directions to reduce its vibration.

The compressors are mechanically and thermally fixed to the outer
shell of the cryostat. 
Hence the heat from the compressors is sunk to the outer shell,
which is designed to work also as a radiator for coolers.

On the other hand, the 2nd-stage cold heads of the coolers are thermally connected to the IVCS,
and the coolers thereby reduce the heat load to the helium tank as discussed above.
Another important role of the cryocoolers is to cool the telescope and IRC to 
a 30~K-level after liquid helium exhaustion, so that we can continue
observations by IRC in the near-infrared.

Two sets of cryocoolers are used for AKARI for redundancy. 
Each set of the coolers has the cooling capacity of 200 mW at 20~K
with the input power of about 100~W. Hence one is enough to
meet the cooling requirements.
In the nominal operation, however, two sets are operated simultaneously
with 50~W input for each set.
If one of the cryocoolers fails on orbit, we will operate
the other cryocooler with the full power (100~W input).

We conducted 40,000~hours of lifetime test for a proto
type model of the two-stage Stirling cooler before launch in the laboratory (see also \cite{narasaki_04}).
As a result, we did not find any significant degradation in the performance of the cooler.


\section{Flight operation}

In this section, we outline the flight operation of the AKARI cryogenic system.
We discuss pre-flight operation, launching operation, and on-orbit operation until
the nominal observation phase started. 

\subsection{Pre-flight operation}

\begin{figure}
  \begin{center}
    \FigureFile(130mm,100mm){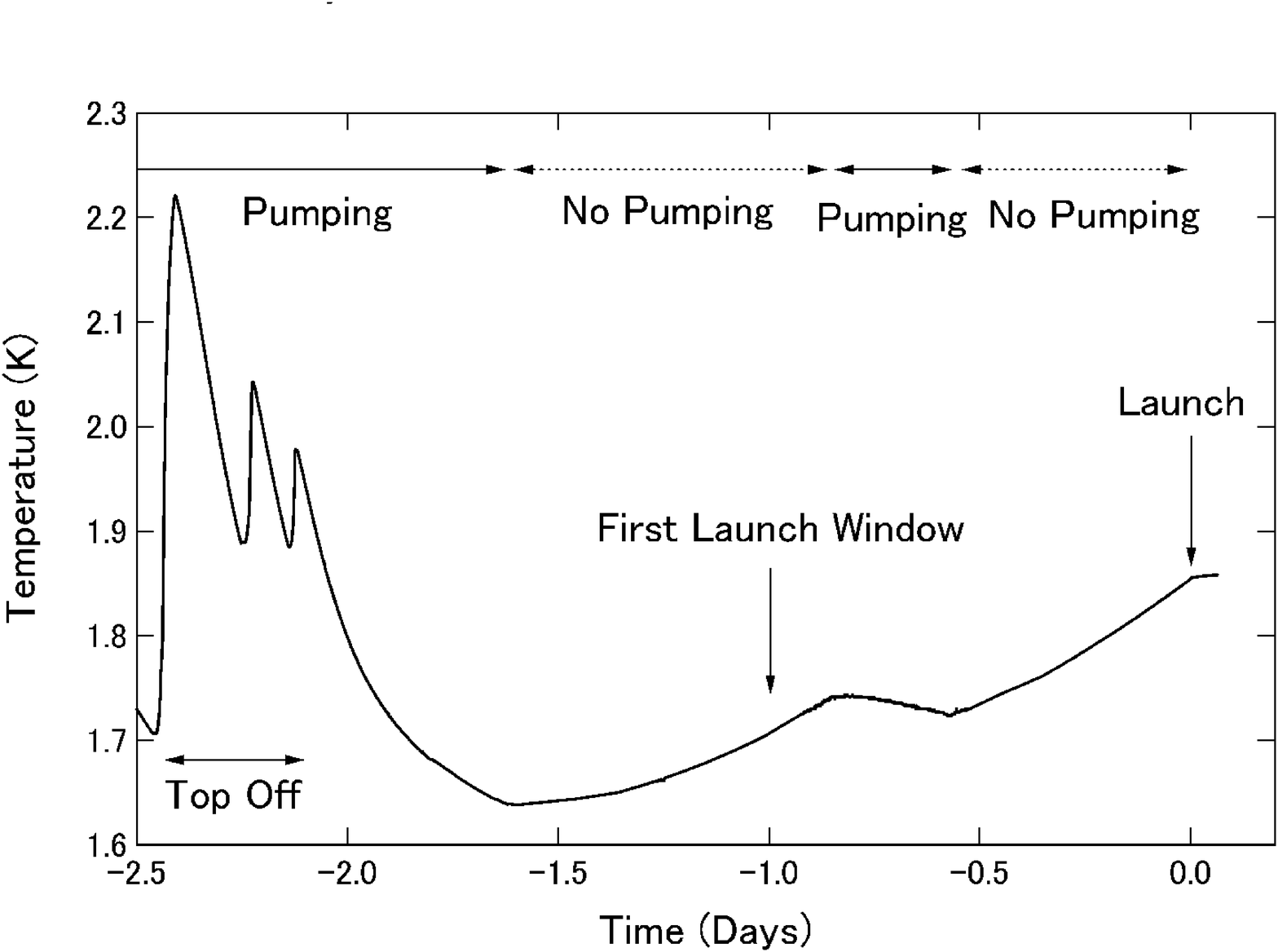}
  \end{center}
  \caption{Temperature change during pre-launch operation.}
\label{fig:pre-launch}
\end{figure}

The most important task during the pre-flight operation
is to fill the helium tank with super-fluid liquid helium (HeII).
If the tank is filled with normal liquid helium (HeI) and 
liquid helium is pumped down to 1.8~K, only 70~\% of the
whole volume of helium tank ($V_{\rm Tank}$) can be filled with HeII.
Hence transfer of HeII, so-called "top-off", is required just before 
the launch.
A series of final transfer of HeII (top-off) to the helium 
tank was performed at the launching pad.
Figure~\ref{fig:pre-launch} shows the temperature change
during this pre-launch operation phase.

We performed three transfers during the low
temperature top-off operation.
The third transfer filled the whole volume (188.7 L) of
the helium tank with HeII at the temperature of 1.978~K.
 
After the third transfer, we pumped down helium for 13.5 hours, and liquid helium 
cooled
from 1.978~K to 1.638~K. 
We estimated that the tank was filled with 180 L (95.4~\% of $V_{\rm Tank}$)
of liquid helium at this point.
Then we stopped pumping, closed all valves, and waited for the
launch. We tried the first launch on February 20, but the launch was called off
because of bad weather.

After the call off, we restarted pumping. Just before we restarted pumping,
the temperature of helium was 1.742~K, because the 
venting valve had been closed for 16~hours.
For this 2nd-day operation, we pumped helium about 6 hours, and the temperature 
of helium went down to 1.725~K.
Then we stopped pumping, and closed all the valves.
We waited about 14 hours before the launch
as shown in figure~\ref{fig:pre-launch}.
We estimate that we had 179~L (26.0~kg or 94.9~\% of $V_{\rm Tank}$) of liquid helium at the temperature
of 1.854~K (significantly below the $\lambda$ point of 2.17~K required) at the time of launch.
The requirements are the volume of 170 L and the temperature 
significantly below the $\lambda$ point of 2.17~K, and they were satisfied at the launch.

\subsection{Lift off}

\begin{table}
\caption{Sequence of events concerning the cryogenic system.}
\label{tab:soe}
\begin{center}
\begin{tabular}{llrl}
\hline
\hline
Date & Time & \multicolumn{1}{c}{Time} & Events \\
(UT) & (UT) & \multicolumn{1}{c}{after} & \\
	& & \multicolumn{1}{c}{launch} & \\
\hline
Feb.21  & 21:28:00 	& 0 s & Lift Off \\
		& 21:31:50  & 230 s & He vent valve opened \\
	    & 21:34:40	& 400 s & He fill valve opened \\
Feb.27  & 11:04:32 	& 5.57 days & Continuous operation of \\
		&			&		& cryocoolers started\footnotemark[$*$]\\
Mar.01  & 07:08:26 	& 7.40 days & Porous plug L closed \\
Mar.14  & 07:35:09 	& 20.42 days & He mass measurement \#1 \\	
Mar.30  & 14:34:43 	& 36.71 days & Cryocoolers stopped\footnotemark[$\dagger$]\\
Mar.31  & 08:11:33 	& 37.45 days & Cryocoolers restarted\footnotemark[$\dagger$]\\
Apr.13  & 07:55:11 	& 50.44 days & Aperture lid jettisoned \\
May 09  & 21:35:55 	& 77.01 days & He mass measurement \#2 \\	
Jul.06  & 21:22:49 	& 135.00 days & He mass measurement \#3 \\	
\hline
    \end{tabular}
  \end{center}
\footnotemark[$*$] The cryocoolers had been operated even before this timing but only intermittently. \\
\footnotemark[$\dagger$] The cryocoolers stopped during this period due to 
unexpected behavior of the power line of the satellite.
\end{table}

\begin{figure}
  \begin{center}
    \FigureFile(90mm,120mm){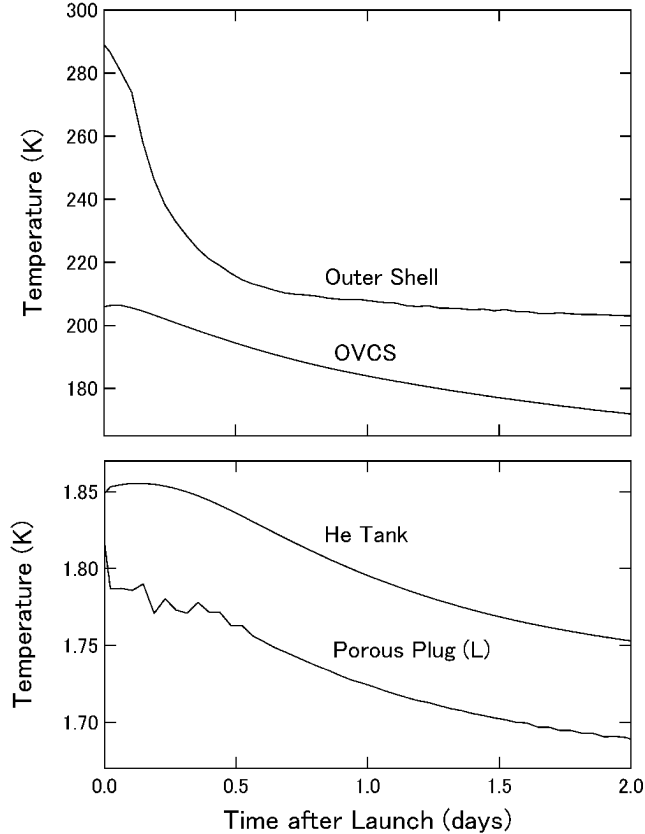}
  \end{center}
  \caption{Temperature change after the launch}
\label{fig:liftoff}
\end{figure}

The AKARI satellite was launched at 21:28 on 2006 February 21 (UT)
from the Uchinoura Space Center by a M-V rocket.
Table~\ref{tab:soe} summarizes sequence of events concerning the 
cryogenics.

At the lift-off, all the valves were closed and the cryocoolers were turned off.
Temperatures and status concerning the cryogenics were
monitored during the launching operation.
The upper panel of figure~\ref{fig:liftoff} shows the temperature change of the outer shell and the OVCS
after the launch. The outer shell started to cool just after the launch,
and went below 220 K within half a day after the launch.
This quick change of the temperature is attributed to the
effective radiative cooling of the outer shell.
The lowered temperature of the outer shell reduced the heat load
inside the cryostat, and thus the temperature of OVCS decreased gradually
after the launch.

The venting valve V6 (figure~\ref{fig:overview}) was opened
230 seconds after the launch, and the
boil-off helium gas started to be vented into space.
The timing to open the valve V6 was determined, so that the ambient pressure 
was low enough for boil-off gas to be vented outside the cryostat, while the acceleration that 
retain the helium toward
the bottom of the tank still remained.
 
The lower panel of figure~\ref{fig:liftoff} shows the temperatures 
of the helium tank ($T_{\rm He}$) and
the down stream of a large porous plug (PP-L) ($T_{\rm PP-L}$).
Before the launch, when the venting valve V6 was closed, 
the temperature of the down stream of PP-L ($T_{\rm PP-L}$) was the same with
that of the helium tank ($T_{\rm He}$).
On the other hand, immediately after V6 was opened (230s after the launch), 
$T_{\rm PP-L}$ dropped about 0.07~K,
which indicated that the porous plug started to work as a phase separator.
The small porous plug (PP-S) shows similar behavior.

During a few hours after the launch, the boil-off gas flow rate was so high 
that the helium tank temperature continued to rise for a while.
The temperature of the 
helium tank reached its maximum of 1.86~K about three hours after the launch,
and started to go down as OVCS started to be cooled.
Oscillation of $T_{\rm PP-L}$ was observed till about half a day after the launch.

As shown above, the helium tank temperature was always well below the $\lambda$ point of 2.17~K.

\subsection{Switching porous plugs}

\begin{figure}
  \begin{center}
   \FigureFile(90mm,120mm){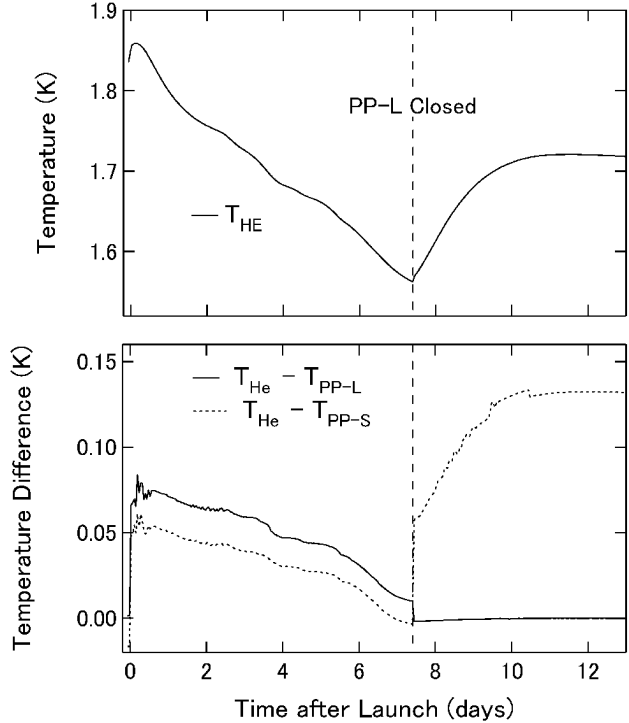}
  \end{center}
  \caption{Temperature change concerning the operation of porous-plugs.}
\label{fig:ppswitch}
\end{figure}

As discussed above, the AKARI cryogenic system has two
porous plugs, PP-L and PP-S, which are optimized for 
flow rates different by an order of magnitude.
When the boil-off gas flow rate was still high just after the launch, we used both PP-L
and PP-S so that we could keep the helium-tank temperature low enough even with high gas flow rate.
But, when the boil-off gas flow rate became low enough that
gas venting only through PP-S is enough, we closed
the valve V5 to disable PP-L.

Figure~\ref{fig:ppswitch} shows the temperature change
of the helium tank ($T_{\rm He}$)
after the launch 
together with the temperature
differences between the downstream of two porous plugs and
the helium tank ($T_{\rm He} - T_{\rm PP-L}$ and $T_{\rm He} - T_{\rm PP-S}$).
After the launch, the temperature differences decreased continuously,
and the temperature difference between the downstream of PP-S and
the helium tank ($T_{\rm He} - T_{\rm PP-S}$) became almost
0 about 7 days after the launch, which indicated that
the operation of PP-S could be unstable.

Hence we closed the valve V5 to disable PP-L 
about 7 days after the launch. 
Figure~\ref{fig:ppswitch} also shows the temperature behavior after
we closed the V5 valve. 
After closing the V5 valve,
the downstream temperature of PP-L rose to the
helium tank temperature, which indicated the termination
of the operation of PP-L.
On the other hand,  
the temperature 
difference between the downstream of PP-S
and the helium tank changed from almost 0~K to 
about 60~mK just after closing V5 valve, which
indicated that PP-S became the dominant venting path.
Since PP-S has higher impedance than PP-L,
the temperature of the liquid helium tank ($T_{\rm He}$) started to
increase gradually after closing the valve.
The temperature of the helium tank
reached its equilibrium temperature (with aperture lid)
of 1.72~K about 3 days after closing the valve.
At the same time,  $T_{\rm He} - T_{\rm PP-S}$
also increased from 60~mK to 130~mK. 

According to the porous-plug component test
before the launch, the temperature difference of
130~mK indicates 
that boil-off rate of helium was 1.67 mg/s.
This 
is consistent with pre-flight tests and prediction of the thermal model of the cryostat.

\subsection{Jettisoning the aperture lid}

Prior to observations, the aperture lid must be jettisoned.
The original plan was to keep the aperture lid on the cryostat for 14 days 
after launch to protect the inside of the cryostat
from contaminants outgassing from the satellite
However, just after launch, 
we found that on-board solar sensors had
problems \citep{murakami_07}.
This forced a delay in the jettison of the aperture lid.
We modified the on-board software to cope with 
solar-sensor problems and then we opened the aperture lid on 2006 April 13, 
51 days after launch. 
As discussed previously, with the aperture lid on, the heat load to the liquid helium is 
three times larger than that with the aperture lid off.
We estimate that  the delay in the aperture lid jettison of 37 days results in a loss of 
100 days of observation time.



\begin{figure}
  \begin{center}
 \FigureFile(90mm,120mm){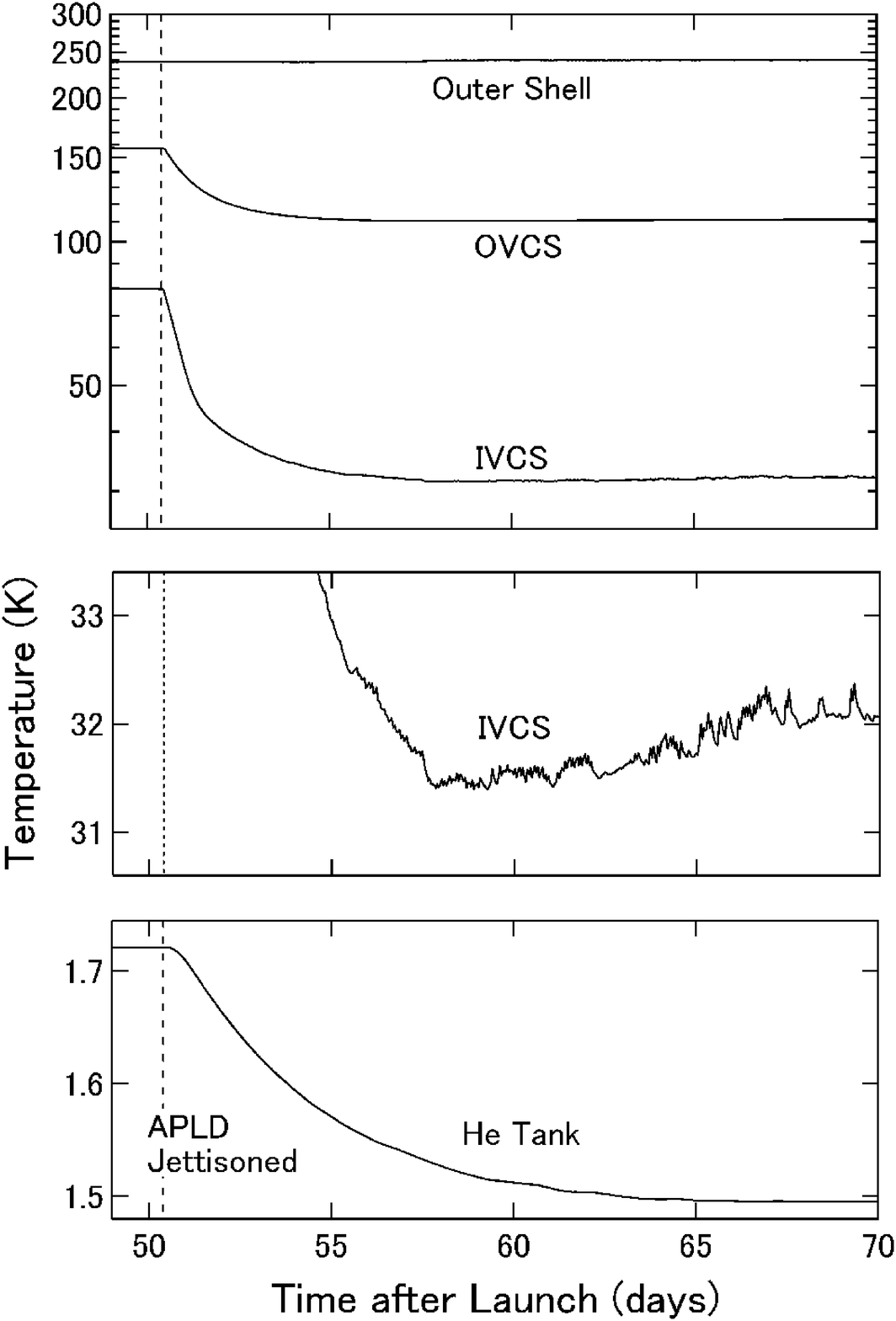}
  \end{center}
  \caption{Temperature change after jettisoning the aperture lid (APLD).}
\label{fig:alid}
\end{figure}

Figure~\ref{fig:alid} shows the temperature change
after jettisoning the aperture lid.
All the temperatures concerning the AKARI cryostat started 
to decrease 
after the aperture lid was jettisoned.
OVCS and IVCS started to be cooled, because they had radiators
(figure~\ref{fig:overview}), 
which had been blocked by the aperture lid
and the radiators started to work once the aperture lid was jettisoned.
Please note that the IVCS temperature reached
its minimum temperature about a week after the aperture lid
was jettisoned. This was caused by 
the balance between the decreasing temperature
of OVCS and the boil-off helium gas flow rate,
which was still higher than its equilibrium value
at this point. After this point, the temperature of 
IVCS started to rise, since the helium gas flow rate decreased continuously. 

As the vapor
cooling shields cooled, the helium tank also started to 
cool gradually. It took about two weeks for the helium tank
to reach its equilibrium temperature of 1.5~K.

\section{Flight performance of the cryogenic system}

\subsection{Comparison with model calculation}

The cryogenic system of AKARI reached its equilibrium state about
two weeks after jettisoning the aperture lid, and we checked
its performance on orbit during the equilibrate phase.

Table~\ref{tab:performance} shows the comparison of the flight performance of the AKARI cryogenics
with the model prediction \citep{hirabayashi_07}. Since the temperatures of various stages 
change on orbit due to the
change of the heat input from the earth, we take the average values (one week in May 2006) on orbit.
As table~\ref{tab:performance} shows, the flight performance 
is quite consistent with the model prediction made before launch.
One can notice small difference in the temperature of the helium tank (flight 
value is lower) and in the temperature of the telescope (flight value is higher). 
This tendency implies that the helium gas flow rate is slightly lower than
the model prediction.

\begin{table}
\caption{Comparison of flight performance with model predictions.}
\label{tab:performance}
\begin{center}
\begin{tabular}{lrrrr}
\hline
\hline
& \multicolumn{3}{c}{Measured Values on Orbit} & Model \\
&	Max	& Min & Average	&  \\
\hline
He Tank (K)	& 1.504	& 1.499	& 1.502	& 1.60 \\
Telescope (K) &	6.575 &	5.874 & 6.175 &	5.6 \\
Cryocooler (K) &	26.10 &	24.45 &	24.92 &	23.9 \\
IVCS(K)	& 34.31	& 31.62 &	32.36 &	24.6 \\
OVCS(K)	& 110.8	& 110.6	& 110.6 &	102.0 \\
Outer Shell (K)	& 244.0	& 241.9	& 243.0	& 238.9 \\
 Compressor (K)	& 248.2	& 246.3	& 247.3	& 240.2 \\
 Cryocooler Power (W) &	100.2	& 98.9	& 99.8	& 100 \\
\hline
    \end{tabular}
  \end{center}
\end{table}

\subsection{Helium mass measurement}

\begin{figure}
  \begin{center}
   \FigureFile(90mm,120mm){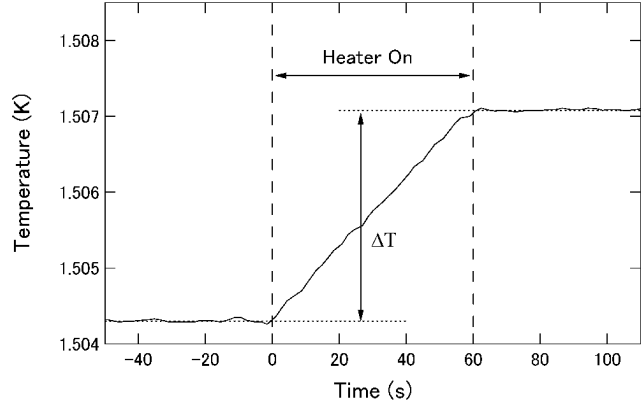}
  \end{center}
  \caption{Temperature profile during the measurement of liquid helium mass. 
We turned on a heater for one minute and observed the temperature increase
due to the additional heat load.}
\label{fig:he_mass}
\end{figure}

We measured the mass of liquid helium by applying
the constant heat load (781~mW) to the helium tank
for one minute, which makes the total heat of 46.86~J.
Figure~\ref{fig:he_mass} showed an example of the temperature profile
during one of the measurements on orbit.
The temperature showed linear increase when the input heater was on,
and became quickly stable just after the heater was turned off.

We made the measurement of helium mass three times so far as shown in table~\ref{tab:he_mass}.
The first one was made before the aperture lid was jettisoned and still the boil-off
rate of helium gas was higher than the nominal value.
The second and the third measurements were performed after the aperture lid was jettisoned and 
when the helium boil-off rate reached its equilibrium.

\begin{table}
\caption{Measurements of helium mass.}
\label{tab:he_mass}
\begin{center}
\begin{tabular}{lrccl}
\hline
\hline
 & Days after & $\Delta T$  & He mass & comments \\
  & launch & (mK) & (kg) & \\
\hline
\#1 & 20.42 & 1.08 $\pm$ 0.03 & 20.95 $^{+0.21} _{-0.31}$ & With aperture lid \\
\#2 & 77.01 & 2.78 $\pm$ 0.03 & 14.71 $^{+0.15} _{-0.22}$ & Without aperture lid \\
\#3 & 135.00 & 3.14 $\pm$ 0.03 & 13.10 $^{+0.12} _{-0.19}$ & Without aperture lid \\
\hline
    \end{tabular}
  \end{center}
\end{table}

From the difference of the second and the third measurements, we estimated
that helium gas boil-off rate was 0.32 $\pm$0.07 mg/s, which
is slightly smaller than the model prediction (0.49 mg/s). This tendency 
is also consistent with the tendency mentioned in the previous subsection on
the comparison between the observed temperatures and model predictions.

\subsection{Comparison with other satellites}

\begin{table}
\caption{Comparison of cryogenic systems for space infrared missions.}
\label{tab:comparison}
\begin{center}
\begin{tabular}{lccrrrrrrl}
\hline
\hline
 & He flow rate & $T_{\rm He}$ & \multicolumn{1}{c}{$P_{\rm He}$} & 
\multicolumn{1}{c}{$V_{\rm Tank}$} & \multicolumn{1}{c}{$T_{\rm Shell}$}  & 
\multicolumn{1}{c}{$H$} & \multicolumn{1}{c}{$P_{\rm He}/H$}  & 
\multicolumn{1}{c}{He hold time} & Ref. \\
  &  \multicolumn{1}{c}{(mg/s)} & (K) &  \multicolumn{1}{c}{(mW)} & 
\multicolumn{1}{c}{(L)} & \multicolumn{1}{c}{(K)} & \multicolumn{1}{c}{(W)} & &
\multicolumn{1}{c}{(days)}\\
\hline
IRAS &  2.4 & 1.80 & 55.4 & 545 & 197 & 342 & $1.6 \times 10^{-4}$ & 302 & (1), (2) \\
IRTS & 4.0 & 1.91 & 78.9 & 100 & 300 & 593 &  $1.3 \times 10^{-4}$ & 37 & (3) \\
ISO & 4.1 & 1.74 & 93.6 & 2342 & 113 & 97.9 & $9.6 \times 10^{-4}$ & 873 & (2), (4) \\
SPITZER  & 0.29 & 1.24 & 5.7 & 360 & 34 & 0.230 & $ 2.5 \times 10^{-2}$ & 1930\footnotemark[$*$] & (5), (6)\\
AKARI & 0.32 & 1.50 & 7.2 & 189 & 243 & 168 & $4.3 \times 10^{-5}$ & 590\footnotemark[$*$] & This work \\ 
\hline
    \end{tabular}
  \end{center}
\footnotemark[$*$] Expected hold time.\\
Reference: (1) \citet{urbach_84}, (2) \citet{holmes_02}, 
  (3) \citet{fujii_96}, (4) \citet{seidel_99}, (5) \citet{finley_04}, and (6) \citet{werner_04}. 
\end{table}

Table~\ref{tab:comparison} shows the comparison of the on-orbit performance of the cryogenic 
systems of representative infrared astronomical satellites.
The second column of table~\ref{tab:comparison} show the boil-off helium gas flow rate ($\dot{m}$) 
in the
equilibrium state on the orbit, the third column shows the heat load to the cryogen ($P_{\rm He}$) 
converted from $\dot{m}$.
AKARI's performance (in terms of the heat load to the cryogen) is much better than those of IRAS, 
ISO and IRTS and is close to that of SPITZER.

However, each cryostat in Table~\ref{tab:comparison} has a
different temperature of the outer shell ($T_{\rm shell}$),
and the radiative heat load on cryogen is different.
To compare the performance of cryostats with
different $T_{\rm shell}$,
\citet{holmes_02} introduces a parameter $H$,
to which the radiative heat load on the cryogen is proportional.
\begin{equation}
H = \sigma_{\rm B} A_{\rm Tank} ( T_{\rm shell}^4 - T_{\rm He}^4), 
\end{equation}
where $\sigma_{\rm B}$ is the Stefan Boltzmann constant and 
$A_{\rm Tank}$ is the surface area of the cryogenic tank.
Here we take $A_{\rm Tank} = S V_{\rm Tank}^{2/3}$, 
and $S$ is a geometrical factor 
for different shape of tanks.
We use $S=6$ as a representative value for cylindrical tanks \citep{holmes_02}.
The 7th column in table~\ref{tab:comparison} shows the parameter $H$ 
and the 8th column shows the parameter ratio $P_{\rm He}/H$, which 
compares the performance of cryostats with different $T_{\rm shell}$.

The ratio $P_{\rm He}/H$ in table~\ref{tab:comparison} shows that IRAS and IRTS have similar performance,
which is slightly better than that of ISO,  
although their sizes are completely different. 
SPITZER has a very low helium gas flow rate and small $P_{\rm He}$.
However, according to the comparisons based on  $P_{\rm He}/H$,
SPITZER does not show very good performance. 
The good performance of SPITZER in $P_{\rm He}$ is mostly due to 
its low $T_{\rm Shell}$.

On the other hand, AKARI shows the best performance in $P_{\rm He}/H$
among the systems listed in table~\ref{tab:comparison}.
Although AKARI has relatively high outer-shell temperature, 
its $P_{\rm He}$ is close to that of SPITZER, whose $T_{\rm Shell}$ is much lower than that of AKARI.
This is mainly due to the good 
efficiency of the two-stage Stirling coolers incorporated in the  hybrid cryogenic design of AKARI.

The 9th column in table~\ref{tab:comparison} shows the on-orbit hold time of liquid helium. 
Although the delay 
in the aperture lid jettison resulted in significant loss of liquid helium,
the expected on-orbit hold time of liquid helium for AKARI is longer than 500 days, which
meets the requirements discussed in section~\ref{sec:req}. 

\section{Orbital change due to vented gas}

\begin{figure}
  \begin{center}
    \FigureFile(120mm,120mm){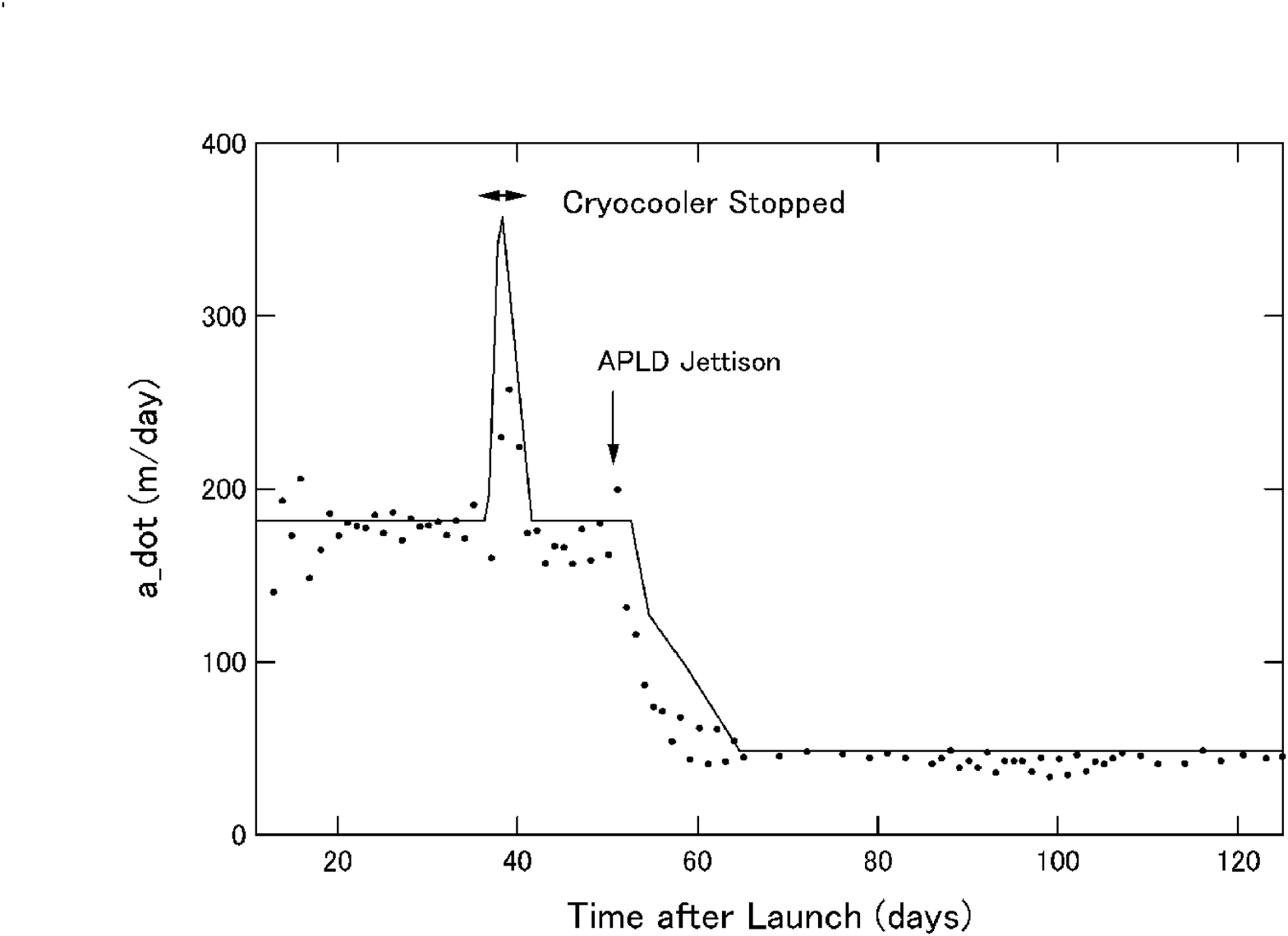}
  \end{center}
  \caption{Rate of change of the semi major axis ($\dot{a}$) for the orbit of AKARI. Dots show observed
values, while the line indicates values obtained by the equation~4
together with $\dot{m}$ estimated on the basis of the cryogenic model \citep{hirabayashi_07}.}
\label{fig:orbit}
\end{figure}

\begin{figure}
  \begin{center}
    \FigureFile(140mm,100mm){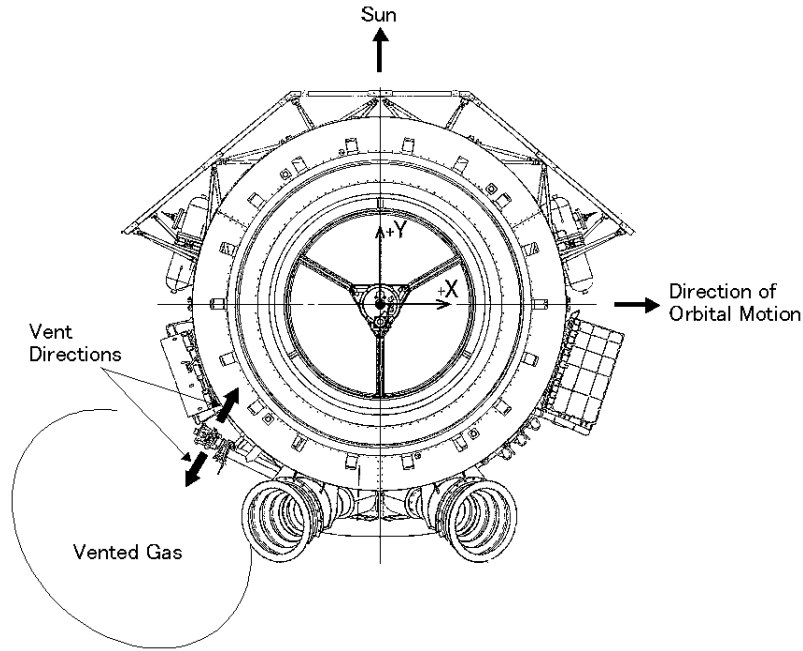}
  \end{center}
  \caption{Top view of the AKARI cryostat and the directions of gas vent.}\label{fig:topview}
\end{figure}

Soon after the launch, we found that the the AKARI orbit was not descending but ascending day by day. 
In response to several operations of 
the cryogenic system, its ascending rate changed quite a lot as shown
in figure~\ref{fig:orbit}. This implies that the orbital change has something to do with helium gas
in the system.

We check the design of the vent path of helium in the AKARI cryogenic system.
Vented gas goes to two opposite directions
from the venting valves as shown in figure~\ref{fig:topview}, 
and thereby the total momentum of the flowing helium gas is canceled.
However,
the pressure due to the thermal motion of the vented gas can push the satellite to one direction,
since the gas is predominantly on one side of the satellite as shown in figure~\ref{fig:topview}.
The venting valves are located at the angle of 45~degrees from 
-Y axis to the -X axis as shown in figure~\ref{fig:topview}.
Hence the gas can push the satellite both to the +X and +Y directions.
Among the two components, 
$+$Y vector works to change the inclination of the orbit, but
its effect is canceled when averaged in one revolution of the orbit. 
On the other hand, the other component, 
$+$X vector, pushes the satellite toward the direction
of the orbital motion of the satellite and hence this can be the $\Delta V$ component to 
increase the orbit.



In the following, we make a rough estimate of if this effect can explain the observed
orbital change.
Small orbital change ($\Delta r \ll r$) 
due to integrated $\Delta v$ can be written as follows
\citep{wertz_99}: 
\begin{equation}
\Delta v = \frac{\Delta r}{2 r}\sqrt{\frac{G M_E}{r}},
\end{equation}
where $r$ is the radius of the orbit (here we assume a circular orbit for simplicity) and 
$M_{\rm E}$ is the mass of the earth.

If we assume that the gas is in thermal equilibrium, we can write the 
average gas thermal velocity $v_{\rm th}$ as a function of temperature as follows:
\begin{equation}
\frac{3}{2} k T = \frac{1}{2} m v_{\rm th}^2.
\end{equation}

Here we assume that the gas motion is completely isotropic, and
only one third of the total energy will contribute to pushing the satellite, 
then the conservation of the momentum gives us the following,
\begin{equation}
M_{\rm sat} \Delta v_{\rm day} = \frac{2}{\sqrt{2}\sqrt{3}} \dot{m} t_{\rm day} v_{\rm th} \eta, 
\end{equation}
where       
$M_{\rm sat}$ is  the mass of satellite,
$\Delta v_{\rm day}$ is $\Delta v$ integrated in one day, 
$\dot{m}$ is helium gas flow rate, 
$t_{\rm day}$ is the length of one day, and
$\eta$ is the efficiency of this effect.

We use $T=$240 K (outer shell temperature) and $r$ = 700 km, 
and we get the following relation between $\dot{r}$ and $\dot{m}$:
\begin{equation}
\dot{r} = 49 \left( \frac{\eta}{0.8} \right)
\left( \frac{\dot{m}}{0.32~\rm{mg/s}} \right)~{\rm m/day} .
\end{equation}
This $\dot{r}$ (circular orbit) can be compared with the observed
$\dot{a}$ (elliptical orbit), since AKARI's orbit is very close to circular.
Figure~\ref{fig:orbit} also  shows a theoretical curve which is derived by using
$\dot{m}$ estimated from the cryogenic model with $\eta=0.8$. 

The observed orbital
change can be reasonably explained by the model prediction.
Hence we conclude that the orbital change is due to the effect that thermal 
pressure of the vented gas is pushing the satellite.

\section{Summary}

The thermal behavior of the AKARI 
cryogenic system has been presented.
The AKARI cryogenic system employs a hybrid system, consisting of
cryogen (liquid helium) and mechanical coolers (2-stage Stirling coolers). 
With the help of the mechanical coolers, 
179 L of liquid helium can keep the instruments cryogenically cooled for more than 500 days. 
The on-orbit performance of the AKARI cryogenics is consistent with the design and
pre-flight test, and is much better than those of previous missions.
Although the cryostat vent was designed to prevent thrust from the venting helium gas, a very small 
thrust from the venting helium is observed anyway. This effect is reasonably explained by the 
thermal pressure of the helium
gas exerted at the vent site which is on the only one side of the space craft.

\section*{Acknowledgements}

AKARI is a JAXA project with the participation of ESA.
We are deeply grateful to all the members of the AKARI team
for their continuous support.
We would  like to cordially thank the M-V rocket team
for their help during the launching operation.
This work was supported in part by a Grant-in-Aid for Scientific Research from
Japan Society for the Promotion of Science (No. 15204013).







\end{document}